# Object-Oriented Translation for Programmable Relational System (DRAFT)

Evgeniy A. Grigoriev,
HTTP://RXO-PROJECT.COM
GRIGORIEV.E@GMAIL.COM
AUTHOR'S ADDRESSES: VOLOTSKOY PER., 13-2-415, MOSCOW, RUSSIAN FEDERATION.

The paper introduces the principles of object-oriented translation for target machine which provides executing the sequences of elementary operations on persistent data presented as a set of relations (programmable relational system). The language of this target machine bases on formal operations of relational data model. An approach is given to convert both the description of complex object-oriented data structures and operations on these data, into a description of relational structures and operations on them. The proposed approach makes possible to extend the target relational language with commands allowing data be described as a set of complex persistent objects of different classes. Object views are introduced which allow relational operations be applied to the data of complex objects. It is shown that any operation and method can be executed on any group of the objects without explicit and implicit iterators. Binding of both attributes and methods with their polymorphic implementations are discussed. Classes can be co-used with relations as scalar domains, in referential integrity constraints and in data query operations.

Categories and Subject Descriptors: F1.1 [COMPUTATION BY ABSTRACT DEVICES]: Models of Computation; D3.3 [PROGRAMMING LANGUAGES]: Language Constructs and Features; H2.3 [DATABASE MANAGEMENT]: Languages --- Database (persistent) programming languages; H2.4 [DATABASE MANAGEMENT]: Systems

## 1. INTRODUCTION

Existent object-oriented languages are result of an evolution of programming languages for computers which use addressable memory [Pratt and Zelkowitz. 2001]. Such memory organization is not only in programmable systems. In particular, relational database management system can be considered as virtual computer providing executing the sequence of elementary operations on persistent data presented as a set of relations. We will name such computer "programmable relational system" (PRS). Implemented in PRS, the data structures and elementary operations on the data are described by relational data model [Codd.1970]. The most important property implemented by PRS is data persistence.

We offer approach to creation object-oriented translator for PRS which is target machine in this case (further: relational target machine). Source language of the translator can be described as declarative object-oriented language. Its commands allow the next. Classes of unique objects having complex (non 1NF) structures can be created. References are allowed. Constraints can be defined on the classes. The objects encapsulate persistent state and behavior. Procedural extensions are used to implement the class methods. The object properties can be redefined in multiple class inheritance. Persistent objects can be created and manipulated by both data access commands and methods. The objects data can be received used ad-hoc queries.

It may be interesting that all operations on complex objects including object methods execution can be performed on any set of the objects without both explicit (described by source language) and implicit (described by target machine language) iterators.

Let us note that command translation is used which differs from program translation as it is usually implied when a object-oriented translation is being discussed. The main difference is that program translation is a process precedes the program execution. As a result, all possible program data structures and procedures on the data are fixed by program code and cannot be changed during execution of the program. Command translation is continuous process concurring with system life. It makes possible to change data descriptions and system functionality. To perform command translation persistent symbol tables have to be used which contain all object metadata and can be considered as part of system catalog.

Our approach allows existing relational DBMSs be extended in the direction set by the "Third-Generation Database System Manifesto"[Stonebraker at all.1990] but in different way that offered by current SQL standards [ISO/IEC 9075:2008. 2008] and existing object-relational DBMSs. Offered interpretation of concept "class" and "object" is similar to the ones used in traditional object-oriented languages [Booch. 1991]. We consider a class as a set of unique persistent objects so tables are not necessary to group the objects and to set them as persistent. Reached object persistence is consequence of total data persistence implemented by relational target machine. Relation target machine also allows the situation be avoided when source object-oriented language includes some awkward features dictated by target machine with addressable memory. For example extents and inverse references used in ODMG data model [Cattell at all. 2000] are explainable only by usage of such target machine where one-direction pointers are the only way to get access to the data stored in the memory.

All data described as complex objects of different classes are presented in form of normal relations (object views) which are calculated according to names and name sequences set by the class description. The object views keep the meaning of the complex objects structures inside the names and headers. The object views (i.e. normal form relations) are the only way to operate with the data of complex objects; therefore no other formal data model than the relational one is necessary to operate with the data. Actually the classes and their implementations are orthogonal to relational data model as it required by [Date and Darwen.1998].

Offered approach has implemented in prototype described in [Grigoriev.2011, 2012].

We proceed as follow. In section 2, the PRS definition of is given and its control commands are listed. These commands are used further to write the results of translation. In section 3, common properties of classes and objects, implemented by offered approach are formulated according to features of relational target machine. In section 4, source language commands and principles of their translation are described. In section 5, compatibility of classes and relations in the system is discussed. New operation of object progress through class hierarchy is introduced in conclusion part.

## 2. PROGRAMMABLE RELATIONAL SYSTEM

PRS maintains an existence of a relational database. Here the term "relational database" is used in a formal way. According to [Date and Darwen. 1998] a relational database is set of named relational variables **R** (relvars) which is interacted by set of transactions **tr**. The relational variables can be real (stored) or virtual (calculated). A set of relvar **R** definitions together forms the schema of a database. Values of the relvars **R** together form total database value. Transactions **tr** are used to change the database value.

PRS is operated by declarative language which allows procedures be defined, stored and executed. The procedures may take parameters and use local variables. In the paper the next five PRS commands will be used

1) To create and describe relational variable the next command is used:

**CREATE** R($a_1$:$D_i$,… $a_n$:$D_j$) KEY(…$a_i$...) FKEY(…) ON(…);
, where **R** is a name of new relational variable, **a** is an attribute of relation, **D** is a domain of the attribute. PRS is supposed to implement a finite set of domains (i/e/ a base scalar types e.g. **INTEGER**, **FLOAT** etc.) and operations on them. **KEY** is a part defining integrity constraints (keys), **FKEY** (…) **ON** (…) is an optional part defining referential integrity constraints (foreign keys).

Virtual relvars are created with a command binding the described variable name **R** with expression **RValue** which returns a relational value.

**CREATE** R … AS RValue
, where **RValue** expression is one of the next:

- A name **R** of other relational variable.

- A composition f(…RValue…) of relational algebra operations **op** on results of other **RValue** expressions $op_1$(… $op_2$(RValue, $op_3$()…)). Next relational operations on **RValue**s are used further:
  - $R_1 \times R_2$ – Cartesian product,
  - $R_1 \cup R_2$ – union operation,

- $R_1 - R_2$ – difference operation,
- $R_1$ JOIN$_{criteria}$ $R_2$ – join operation (join criteria is in lower index). LEFT JOIN$_{criteria}$ operation is also used.
- $R[a_1, a_2, …]$ – projection operation (a is attribute),
- R WHERE criteria – selection.
- R RENAME a AS b – attribute renaming operation.

This relational operation can contain scalar operations on values of attributes **a**, which are defined for domains **D** of the attributed.

– A procedure which returns the value by **return** RValue operator.

  **begin**
   …
   **return** RValue;
  **end**;

– An explicit relational value given by user.

2) To set values of relational variables the assignment command is used

**SET** R := RValue

If the right part of assignment command contains explicit relational value, such command is considered as data input one.

This command is equal to traditional relational commands INSERT, UPDATE, DELETE [Date and Darwen, 1998] which also are used further.

3) To get values from relational database the next command is used.

**GET** RValue

This command is considered as data output one and can be used as ad-hoc query. It doesn't change the total database value.

4) We consider transactions as sequences of elementary operations changing the database state (i.e. as procedures), either all occur, or nothing occurs. They may contain commands changing database schema or/and values of relational variables. A transaction **tr** can be defined and stored with the next command

**TRANS** tr (parameters_definition)
**AS begin**
 …
 **CREATE** R(…$a_i$:$D_i$...);
 **SET** R := RValue;
 … **end**;

5) To execute the transaction **tr** the next command is used

**EXEC** tr(parameters);

Direct execution of transaction formed of sequence of command is also possible in PRS.

**EXEC begin** … **end**;

These commands **CREATE**, **SET**, **GET**, **TRANS** and **EXEC** are used further to write result of translation.

## 3. CLASSES AND OBJECTS

Our main aim is a language consisting of commands which allow persistent data be described as a set of unique complex objects of different classes. Let us formulate how these common object-oriented concepts may be implemented considering the features of the relational target machine.

In programming languages any complex data object is built as a set of simplest data objects of data types which are realized by target machine [Pratt and Zelkowitz. 2001]. We use this principle to build a complex data object using data types realized by the relational target machine. In this case, an object variable is a set of named components which are relational variables.

(Remark. The term "object components" is equal to "object attributes". We use it in further text to distinguish "object components" from "attribute of a relation".)

Let us note that a relation have two characteristics: arity(A) and cardinality(M). Usually these characteristics are implied to have any values. But object components can be described by simpler language constructions. Such description fixes one or both relation characteristics to unity. For example, components described as a tuple can be considered as a relation variable with cardinality fixed to unity. As a result the next kinds of components are possible: relations (A:n, M:n), sets (A:1, M:n), tuples (A:n, M:1), scalars(A:1, M:1). It is correct to say that the target relation machine allows any object value be described as a set of values which are not more complex than relations.

(Remark. Further only simplest scalar components (simple ones) and most complex relation components (complex ones) are considered for sake of simplicity. The sets and the tuples can be reasoned by analogy.)

Objects also have a behavior which is defined by a set of a methods allowing state of the objects be changed.

The components and the methods together form an object interface. A class is a set of objects having the same interface; this interface is described in class specification. The objects can be accessed as elements of the class. Classes can be inherited. Multiple inheritance is possible. Child class specification is result of UNION set operation on specifications of parent classes and the set of own components and methods (virtual inheritance).

A class specification is distinguished from an implementation of the class. The implementations are set separately for each component and method. Class components can be stored or calculated in different ways. Thus persistence of data is the property encapsulated in object components and defined by their implementation. Class methods are implemented by procedures. All the implementations can be redefined during inheritance.

Each object is unique among other ones. This common object property is implemented in relational target machine with OID values of dOID domain. The OID value is not associated with the object state. It is generated by the RPS at object creation and stays unchanged during all object life. The OID values are not accessible for users.

All OIDs of objects of some class are united in corresponding reference type having the same name. Next operations are defined for reference variables: assignment, comparison and dereferencing operation. Reference types are scalar ones which the set of domains D is being expanded by.

Besides the OID, the objects uniqueness inside their classes can be defined by optional explicit keys. With them, explicit foreign keys are possible too.

## 4. SOURCE LANGUAGE COMMANDS AND TRANSLATION

This part describes main commands and expressions of the source language and gives principles of their translation into machine language commands. Further the source language commands and the expressions are discussed in next order:
- Class creation command,
- Data access commands,
- Implementation command (implementing expression),
- Implementation command (binding expression),
- Object creation command.

For each of the listed group of commands, their essence is described firstly. Then the principles of their translation are given. Examples are used when necessary. The syntax of source language used in the examples is close to the SQL and to the one realized in prototype described in [Grigoriev, 2011].

Let us start with a common example which uses commands listed above. a simple model of a trade company is created in the example. Classes are specified and implemented and objects are created. Data structure created by these command is shown. Data access operations (inc. queries) are given. Then the classes are inherited and re-implemented and objects of child classes are created too. Group method execution and data access operations to polymorphic classes are illustrated.

## 4.1 Common example

Classes are created and specified by command **CLASS ...** (this command is described in 4.2). The next classes is used as example ones in the paper. Simple class **BANKS** contains data on banks serving contractors.

```
CLASS BANKS
( Name STRING
);
```

Contractors are unique with their IDs. Also each of class **CONTRACTORS** objects references on some **BANKS** object.

```
CLASS CONTRACTORS
( Name STRING,
  Bank BANKS,        //reference
  ID STRING
)KEY(ID);
```

Class **GOODS** describes assortment of sold goods which are unique by their nomenclature articles. Objects of class **GOODS** also contains information on the goods turnover and current quantity of pieces on stock.

```
CLASS GOODS
( Art STRING;
  Turnover SET OF    //complex component
   ( DocN STRING,
     Cntr CONTRACTORS,     //reference
     Pieces INTEGER
   )KEY(DocN),             //component key
   Pieces INTEGER    //...remain on stock
)KEY(Art);
```

Documents on deliveries and shipments are described by class **DOCS**. Documents are unique with their number **DocN**. This class contains method **DoShip**. The attribute **Art** of complex components **Items** defined as foreign key on key component **Art** of class **GOODS**.

```
CLASS   DOCS                                              //
(E01)
( DocN STRING,
  Date DATETIME,
  Comment STRING,
  Cntr CONTRACTORS,           //reference
  DoShip(inDate DATETIME),    //method
  Items SET OF      //complex component
   ( Art STRING,
     Pieces INTEGER
   )KEY(Art)                  //component key
)KEY(DocN)                    //class key
REFERENCE Items(.Art)
  ON GOODS(.Art)              //foreign key
```

Let us note that ad-hoc queries on classes can be executed right after the classes are defined (data access commands are described in 4.3) f.e. the next query.

```
SELECT #S.DocN,
       #S.Comment,
       #S.Items.Art,
       #S.Items.Oty
FROM DOCS<DocN LIKE "%1"> #S; // #S is alias
```

In this point classes is not implemented so result of the query is empty. Class must be implemented fully to allow objects be created. Some class components can be implemented as stored (implementation command is described in 4.4).

```
ALTER BANKS REALIZE Name AS STORED;

ALTER CONTRACTORS REALIZE Name, Bank, ID AS STORED;

ALTER GOODS REALIZE Art AS STORED;

ALTER DOCS REALIZE DocN, Date, Comment, Cntr, Items AS STORED;
```

Other class components can be implemented as calculated. E.g. complex component **Turnover** is implemented in class **GOODS** as calculated by a query on data presented in class **DOCS**

```
ALTER GOODS REALIZE Turnover AS
SELECT #g.DocN,
       #g.Cntr,
       SUM(#g.Items.Pieces) AS Pieces
FROM DOCS #g
WHERE #g.Items.Art = Art
GROUP BY
  #g.DocN,
  #g.Cntr;
```

Here, in comparition `#g.Items.Art = Art` last **Art** is the name of simple component of class **GOODS**. Simple component **Pieces** is implemented in class **GOODS** by the next procedure

```
ALTER GOODS REALIZE Pieces
AS {
  DECLARE tmpPieces INTEGER;
  tmpPieces :=
  SELECT SUM(#g.Items.Pieces) AS Pieces
    FROM DOCS #g
    WHERE #g.Items.Art = Art;
  IF(tmpPieces IS NULL)
    THEN tmpPieces := 0;
  RETURN tmpPieces;
}
```

Methods are implemented by a procedure. Method **DoShip** is implemented by the next procedure in class **DOCS**

```
ALTER DOCS REALIZE DoShip(inDate DATETIME)
AS {
  IF(Date IS NULL) THEN
  BEGIN
    Date := inDate;
    Comment := "Shipped!";
  END
}
```

After a class was implemented fully its objects can be created. Command NEW is used to create objects (this command is described in 4.6). In this command a part after **WITH SET** contains constructing expressions.

```
NEW CONTRACTORS WITH SET
  .Name:="TheShop",
  .Bank := (NEW BANKS WITH SET
              .Name:="TheBank"),
  .ID:="CoID001";
```

This command contains nested NEW command; so reference on new-created **BANKS** object is used to initialize the reference component **Bank** of new-created **CONTRACTORS** object. Objects are available as elements of class. In the next command the reference component **Bank** is initialized with reference on existing object using expression **FIRST OF**

**BANKS<.Name="TheBank">** returning reference on the only object of class **BANKS** which satisfies to condition **<.Name="The Bank">**.

```
NEW CONTRACTORS WITH SET
 .Name:="TheRetail",
 .Bank:=(FIRST OF BANKS<.Name="TheBank">),
 .ID:="CoID002";

NEW GOODS WITH SET.Art:="Tie";

NEW GOODS WITH SET .Art:="Axe";

NEW DOCS WITH SET .DocN:="Ship1",
  .Cntr:=FIRST OF CONTRACTORS<.Name="TheShop">;

NEW DOCS WITH SET .DocN:= "Ship2",
  .Cntr:=FIRST OF CONTRACTORS<.ID="CoID001">;

NEW DOCS WITH SET .DocN:="Ship3",
  .Cntr:=FIRST OF CONTRACTORS<.ID="CoID002">);
```

Existing objects can be modified

```
INSERT INTO DOCS<.DocN = "Ship1">.Items (Art, Pieces)
   VALUES ("Axe", 2);

INSERT INTO DOCS<.DocN = "Ship2">.Items (Art, Pieces)
   VALUES ("Axe", 5);

INSERT INTO DOCS<.DocN = "Ship2">.Items (Art, Pieces)
   VALUES ("Tie", 10);
```

Figure 1 shows data structure being result of the given commands execution. Let us note that (according to implementation) the values of components **Turnover** and **Pieces** (underlined) of class **GOODS** objects are calculated from values stored in class **DOCS**.

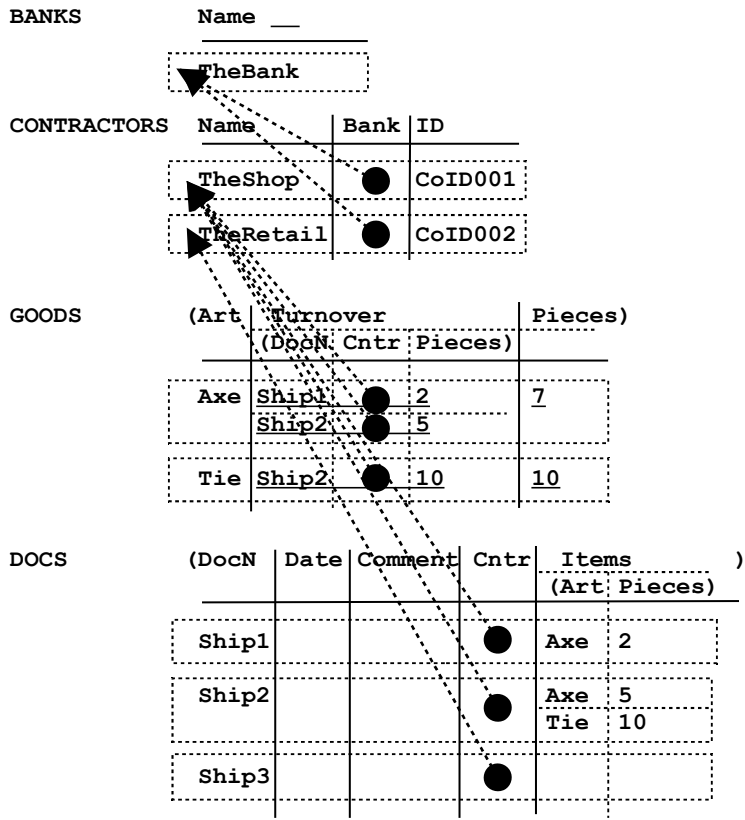

Fig.1. Example object data structure
(Each rectangle contains data on separate object, arrows mean references. Values of calculated components are underlined).

Now the earlier used ad-hoc query

```
SELECT #S.DocN,
       #S.Comment,
       #S.Items.Art,
       #S.Items.Qty
FROM DOCS<DocN LIKE "%1"> #S;
```

returns the next result

```
DocN        Comment         Items       Items
                            .Art        .Pieces
--------------------------------------------------
Ship1                       Axe         2
```

Here expressions "**Items.Art**" and "**Items.Pieces**" are used as a name of relational attribute keeping the meaning of complx structure defined by class specification.

Suppose a class **SALES** inherits both the existing class **DOCS** and new class **VALUERECORDS**. There is complex component **SaledItems** containing data on sold goods in new class **SALES**.

```
CLASS VALUERECORDS
( ...
  Amount FLOAT,...
)...
```

The new component **SaledItems** is implemented as stored too.

**ALTER SALES REALIZE SaledItems AS STORED;**

Implementations of both inherited component **Items** and method **DoShip** are changed in this class. Inherited component **Items** is calculated now.

```
ALTER SALES REALIZE Items AS
  SELECT Art, SUM(Pieces) FROM SaleItems GROUP BY Art;
```

Now inherited method **DoShip()** initializes inherited components of both parent class **DOCS** and **VALUERECORDS**.

```
ALTER SALES REALIZE DoShip(inDate DATETIME)
AS {
  IF(Date IS NULL) THEN
  begin
    Date := inDate;
    Comment:= "Sold!";
      Amount:=SELECT  SUM(#si.Pieces,#si.Price)  FROM  SalesItems #si;
  end
};
```

New object of class **SALES** is created and modified

```
NEW SALES WITH SET
  DocN := "Sale1",
  Cntr:=FIRST OF CONTRAGENTS<.ID="CoID002">;

INSERT INTO SALES<.DocN="Sale1">.SaleItems (Art, Price, Pieces)
  VALUES ("Tie", 10, 30);           //...

INSERT INTO SALES<.DocN="Sale1">.SaleItems (Art, Price, Pieces)
  VALUES ("Tie", 11, 20);           //the same Art but other Price

INSERT INTO SALES<.DocN="Sale1">.SaleItems (Art, Price, Pieces)
  VALUES ("Axe", 20, 50);
```

Let us execute the method for some of parent class objects

```
EXEC DOCS<DocN LIKE "%1">
            .DoShip(...);
```

Figure 2 shows data structure being result of the given commands execution. Let us pay attention to data dependences set by implementation of the classes. Underlined values are calculated ones. In each object of class **SALES**, its component **Items** inherited form class **DOCS** and component **Amount** inherited form class **VALUERECORDS** are calculated from stored component **SaleItems**. The values of components **Turnover** and **Pieces** (underlined) of class **GOODS** objects are calculated from values of components **Items** which is stored in class **DOCS** and calculated in child class **SALES**.

```
GOODS         (Art | Turnover              | Pieces)
                    | (DocN  Cntr  Pieces) |
              ─────────────────────────────────────
              Axe | Ship1   ●    2        | 57
                  | Ship2   ●    5        |
                  | Sale1   ●    50       |
              ─────────────────────────────────────
              Tie | Ship2   ●    10       | 60
                  | Sale1   ●    50       |

DOCS          (DocN | Date | Comment  | Cntr | Items          )
                                              (Art   Pieces)
              ──────────────────────────────────────────────
              Ship1 | ...  | Shipped! |  ●   | Axe    2
              ──────────────────────────────────────────────
              Ship2 |      |          |  ●   | Axe    5
                                              Tie    10
              ──────────────────────────────────────────────
              Ship3 |      |          |  ●   |

VALUERECORDS(                                   Amount)

SALES        (                                           SaleItems
)                                                        (Art  Price
Pieces)                                                           )
              ──────────────────────────────────────────────────────
              Sale1 | ... | Sold! | ● | Axe  50 | 1520 | Axe 20  50
                                        Tie  50          Tie 10  30
                                                         Tie 11  20
```

Fig.2. Example object data structure after all changes. Unchanged classes **BANKS** and **CONTRACTORS** are not present.
(Each rectangle contains data on separate object, dots mean references. Values of calculated components are underlined).

Let us execute the same query again

```
SELECT  #S.DocN,
        #S.Comment,
        #S.Items.Art,
        #S.Items.Oty
FROM DOCS<DocN LIKE "%1"> #S;
```

Now the result of the query is

```
DocN    Comment    Items      Items
                   .Art       .Pieces
---------------------------------------
Ship1   Shipped!   Axe        2
Sale1   Sold!      Tie        50
Sale1   Sold!      Axe        50
```

Column **Comment** shows that proper implementation of method **DoShip(...)** was executed for each of objects. Columns **Items.Pieces** shows other type of polymorphism. Value 2 is stored in complex component **Items** (as it is implemented in class **DOCS**) and the values 50 is result of the component calculation, as it is implemented in class **SALES**. The correct implementations are bound during the query execution (the binding is described in 4.5).

The query is applied to class **DOCS** only. But ability to inherit the class and to redefine its implementations allows data presented in result of the query be stored and/or calculated in different ways. It's possible to expand the data schema with new classes and to get new data by unchanged data access operations over classes.

Let us note that after method **DoShip()** execution the query on abstract class **VALUERECORDS**

**SELECT #vr.Amount FROM VALUERECORDS #vr;**

returns (according to implementations of child class **SALES**) the amount of the posted sale

```
Amount
--------
1520
```

### 4.2 Class Creation Command

The command creates class (i.e. set of objects) and describes its specification.

#### 4.2.1  Description of Class Creation Command

Class creation command has the follow syntax

CLASS $D_{n+1}$ EXTEND $D_n$…
( …
  $^{sc}C_i:D_a$, …
  $^{r}C_j(a_1:D_b, …)$KEY($a_1$…) , …
  $M_k$(…),…
) KEY($^{sc}C_i$…)
REFERENCE (…) ON (…)
, where $D_{n+1}$ – name of new class (and corresponding reference type), EXTEND $D_n$…– optional part listing parent classes; further (in parentheses) class are described with, $^{sc}C_i$ – simple components defined on some $D_a$ domain, $^{r}C_j(a_1:D_b, …)$KEY($a_1$…) – complex components (relation with schema given), $M_k$(…) – method; KEY($^{sc}C_i$…) – optional part describing integrity constrains (class keys consisting of scalar components), REFERENCE (…) ON (…) – optional part describing referential integrity constraints (foreign keys).

#### 4.2.2  Translation of Class Creation Command

During translation of class creation command the translator performs the next action.
- All information about new class (names, structures, links etc.) is analyzed and entered into symbol table
- In relational target machine a set of relations are created which represent objects data.

The objective of class specification command translation is to create a schema of representation of object value structure into data structures possible in target machine (i.e. in relations). Such schema is basis for translation of other commands manipulating objects of the class. The object representation into relation is based on already formulated condition: any object state can be described as set of values which are not more complex than relations. It allows the any object value structure be represented into relations almost directly. The main principle is the next: different components are represented in different relations. But since different kinds of components exist, so the representations schemas are different for the different kinds of components.

− Values of all scalar components $^{sc}C_i$ of all class D objects are represented in single relation $R_D$ (further – class scalar relation)

$(_1OID \times \, _1{}^{sc}C_1 \times _1{}^{sc}C_2 \times \ldots \times _1{}^{sc}C_n) \cup (_2OID \times _2{}^{sc}C_1 \times \ldots \times _2{}^{sc}C_n) \cup (\ldots) \cup \ldots \to R_D$,

where $_jOID$ is identifier of some object, $_j{}^{sc}C_i$ is some scalar component of this object, n is number of scalar components in class D

One class scalar relation $R_D$ conforms to each class D. One and only one tuple of the class scalar relation conforms to each object of the class. OID attribute is mandatory in the scalar relation. Other attributes of the scalar relation conform to class scalar components exactly.

- OID is key attribute.
- If scalar class components form class key, the corresponding attributes form scalar relation key.

− Values of each complex component $^rC_i$ of all class D object are represented in relation $R_{D.Ci}$ (further – class relation of complex component).

$(_1OID \times \, _1{}^rC_1) \cup (_2OID \times _2{}^rC_1) \cup \ldots \to R_{D.C1}$

$(_1OID \times _1{}^rC_2) \cup \ldots \to R_{D.C2}$

…

$(_1OID \times _1{}^rC_m) \cup \ldots \to R_{D.Cm}$

Here $_jOID$ is identifier of some object, $_j{}^rC_i$ is some complex component of this object, m is number of complex components in class D.

Number of the relations $R_{D.Ci}$ is equal to number of set components. OID attribute is mandatory in the relation $R_{D.Ci}$. Other attributes of the relation $R_{D.Ci}$ conform to attributes of corresponding complex component exactly.

- Key of the set relation aggregates OID attribute and the attributes defined as key ones in corresponding component .
- If attributes of set component form global unique constraint on complex component tuples, then corresponding attributes form key of relation $R_{D.i}$.

(Remark. Such constraint is not class key because complex components can be empty.)

− If scalar components or attributes of complex components form a foreign key, then the corresponding attributes of corresponding class relation form foreign key too.
− At that, all references (simple reference components and reference attributes of complex components) are represented in class relations as attributes defined in domain of object identifiers dOID.

Thus all data of all class D objects are represented in one scalar relation $R_T$ and several relations $R_{D.i}$ , named together the class relations. The number of class relation depends on class structure only. At that all constraints defined for class T are represented in constraints set on the class relation. If a class is inherited form other classes its data are represented in both own class relations and class relations of all parent classes.

The described process can be informally defined as relational memory configuration. It's very important that the class relations are not stored but virtual ones. They allows data *be presented* in relation target machine according to class specification. The ways of how the data are calculated to be presented in class relations depend on components implementations only. On this stage of discussion the class relations are accepted as existing and containing data of all objects of the class. Details on class relations calculation will be discussed further (see 4.4 and 4.5).

*Example.* Source command of class **DOCS** creation (E01) is translated into the next sequence of relational target machine commands which create two class relations.

```
EXEC begin
    CREATE R_DOCS( OID: dOID,  DocN:STRING, Date: DATETIME,   Cntr:dOID)
        KEY(OID) KEY(DocN)
        FKEY(Cntr) ON R_CONTRACTORS(OID) AS …;
    CREATE R_DOCS.Items( OID: dOID, Art: STRING,  Pieces:INTEGER)
        KEY(OID, Art)
        FKEY(Art) ON R_GOODS(Art) AS …;
end
```

### 4.3 Data Access Commands

Data access commands are used both to change and to get data described as a set of complex objects. This part devoted mainly to object views which represent data operated by data access commands. The object views are described and then operations which calculate object view are given.

#### 4.3.1   From Paths to Relations

All data access commands use reference path expressions (paths). The paths are name sequences determined by structures and references defined in class specification. Idea of paths - to present hierarchy, to specify a part of a general - seems obvious enough.

Dot notation is used to write a path. Simplest paths include just one name. Type of the path is defined by type of its last element.

Paths used in data access command begin with any expression which means set of objects. In a global context (where commands are executed) any paths begin with a name of a class which are only names defined in global context.

The paths of base scalar types are terminal ones and don't allow continuations.

The paths ended with name of reference or with name of complex component have continuations (further : post-paths). Such paths will be named non-terminal.

*Example.* Path **DOCS.Cntr** (of reference type **CONTRACTORS**) is non-terminal and has next post-paths among other.

**.Name**

**.Bank.Name**

Let us note that post-paths begin with a dot in used source language syntax.

Any name of a class or reference existing in a path can be added with object selection expression.

name_of_class_or_reference<conditions_list>

This expression restricts the set of objects defined by the path using conditions_list. Here each of conditions is applied to scalar post-paths available after name_of_class_or_reference. Conditions can be combined with traditional logical operation AND, OR and NOT. Also new low-priority logical operation is added. Its name is intertuples_AND. Simple comma "," are used to write this operation in source code.

The necessity of the new logical operation is dictated by the fact that complex components are relations containing set of tuples and selection cases inexpressible by usual in-tuple logical operations are possible. The example of such case is given below.

*Example.* To access class **DOCS** objects which contain in attribute **Art** of complex component **Items** both **Tie** and **Axe** string values (in different tuples) next path is used

**DOCS<.Items.Art = "Tie", .Items.Art = "Axe">**

Here the objects have to be selected which contains the tuples in complex component **.Items**, which satisfy to all different conditions applying to the same attribute **.Art**. Traditional in-tuple operation OR cannot be used to combine the conditions because it allows object be selected which satisfy only to one of given conditions. Traditional in-tuple operation AND cannot be used because the scalar tuple attribute cannot be equal to different values simultaneously.

Object selection expression can be nested and combined arbitrarily.

The next is base principle of how the paths are used in data access commands:
Any non-terminal path can be considered as a name of a relation; any scalar post-paths of this path can be considered as names of attributes of this relation. Such relations will be named as O-views (object views) further.

*Example*. Expression **DOCS<.Date >= '2010.01.01'>.Cntr** is non-terminal path which has, among other, post-paths **.Name**, **.ID** and **.Bank.Name**. It means, that the next O-view (i.e. relation) can be used:

```
DOCS<.Date >= '2010.01.01'>.Cntr
   (
   .Name ,
   .ID ,
   .Bank.Name
   )
```

Multitude of other O-views can be used, e.g.

```
BANKS
   ( .Name)

GOODS.Turnover
   ( .Cntr.Name, .DocN , .Date, .Pieces )

GOODS<.Art LIKE " Sometext%">.Turnover
   ( .Cntr.Name, .DocN , .Date, .Pieces)
```

etc. and etc.

Here, the path expression (e.g. **GOODS.Turnover**) is considered as just name of relation i.e. as string identifier which differs from other name (e.g. **GOODS<.Art LIKE "Sometext%">.Turnover**). Also expressions like **.Art** , **.Pieces** **.Turnover.DocN**, **.Turnover.Date** etc. are considered as unique names of the view attributes. All these names keep the meanings of structures which exist in complex objects. They allow the data be presented in form of normal relations, using the names and name sequences defined in class specification. As a result, transition from complex object description to relational representation of the object data is imperceptible for users.

There is no certain relationship between the set of classes and the set of O-views. One class can relate to many O-views. One O-view can present data of different classes (combined by references and/or in inheritance). The set of classes and the set of O-views are linked with common names only. It's possible to say that each of the O-views is not defined separately but the total set of O-views (i.e. relations) is defined by set of classes. In this way all data of complex objects are represented for user simultaneously as a set of normal relations.

### 4.3.2 O-views in Data Access Commands

O-views are relations. Therefore usual relational command can be applied to O-views to access complex object data.

*Example*. Next command changes value of simple components of some objects

```
UPDATE CONTRACTORS<.ID = "..."> SET (.Name := "...");
```

*Example*. Next command adds tuple into complex components of some objects

```
INSERT INTO DOCS<.DocN = "...">.Items (.Art, .Pieces)
   VALUES("Hat", 1);                    //...no such "Art" - error
```

*Example*. Next query returns the data from objects forming complex reference structure

```
SELECT
  #gt.DocN,
  #gt.Cntr.Name,
  #gt.Cntr.Bank.Name
FROM GOODS<.Art LIKE "A%">.Turnover #gt;
```

The result is

```
DocN    Cntr        Cntr
        .Name       .Bank
                    .Name
---------------------------
Ship1   TheShop     TheBank
Ship2   TheShop     TheBank
Sale1   TheRetail   TheBank
```

Let us note that O-views are defined only by class specifications. Therefore if class are defined (this means that class specification are defined too) it is possible to execute commands to access class objects data even if no object of the class exists or the class is not implemented yet. In this way class is considered by users as a set of objects which are accessible as elements of the class. Object references and collections of the references (e.g. class extents) are not obvious to access data of the objects with this ability.

Let us note also, that offered approach allows data be accessed both by references and against references.

*Example*. It's possible to find data of class **CONTRACTORS** objects referenced by defined class **DOCS** objects.

```
SELECT
  #c.Name
  #c.ID
  #c.Bank.Name
FROM DOCS[.DocN LIKE "%1"].Cntr #c;
```

The result is

```
Name        ID          Bank
                        .Name
-----------------------------
TheShop     CoID001     TheBank
TheRetail   CoID002     TheBank
```

Also it's possible to find data of class **DOCS** objects referencing to defined class **CONTRACTORS** objects.

```
SELECT
  #d.DocN
  #d.Items.Art
  #d.Items.Pieces
FROM DOCS[.Cntr.Name = "TheRetail"] #d;
```

The result is

```
DocN        Items       Items
            .Art        .Pieces
-------------------------------
Ship3
Sale1       Axe         50
Sale1       Tie         50
```

### 4.3.3 Translation of Data Access Commands

O-views are not defined manifestly like e.g. SQL views are defined. O-views can be calculated. Any command which manipulates object data contains anyhow some path and its post-paths which together form O-view signature. Generally translation of the data access command includes the next steps
1. The command is parsed to find O-view signatures.
2. Expressions are built which calculate the O-views defined by found signature.
3. In the command, all expressions formed O-view signatures are replaced with expressions calculated the O-views.

Thus the main point of the data access commands translation is O-view calculation. Further set of operations used for O-view calculation is considered. Their base operands are class relations $R_{D...}$ described in previous part. Their results are relations $_oR(OID, …)$. Mandatory OID attribute identifies objects whose data are contained in the relations. Let us note that this attribute is not defined in O-view signature explicitly. Considering that O-views are used in commands only and the results $_oR(OID, …)$ of their calculation aren't accessed directly, the OID attribute is implicit for the users (this is logically for attribute containing system values). The OID attribute is important for system functioning. It allows the structure of the result relations be unified with class relations; so the result relations can be used in further calculation as well as the class ones. Also it's possible to get OID value (i.e. reference on the object) by values of the object components.

#### 4.3.3.1 Simple Projection

If O-view signature is a subset of one of class relations $R_i$ then simple projection

**GET** $R_i$[OID, …]

is used to calculate it.

*Example*. O-view **DOCS.Items(.Art)** is calculated as projection

**GET** $R_{DOCS.Items}$[OID, Art].

#### 4.3.3.2 In-Class Join

Suppose O-view combines data from different class relations of some class D; then result is calculated by LEFT JOIN of class scalar relation on common OID attribute. Also attributes of class relations of complex components $R_{D.C}$ are renamed to fit post-path expressions given in the O-view signature.

**GET** $R_D$ LEFT JOIN$_{OID}$ ($R_{D.C}$ RENAME a as C.a …) …[OID, …, C.a, …]

Required in-class joins perform before in-reference join operations described further.

*Example*. For example O-view

**DOCS(.DocN, .Items.Art)**

is calculated as

**GET** ($R_{DOCS}$ LEFT JOIN$_{OID}$ ($R_{DOCS.Items}$ RENAME Art AS Items.Art))[OID, DocN, Items.Art]

#### 4.3.3.3 On-Reference JOIN in O-View Attributes

Suppose O-view attribute contains reference ref on object of class $_{ref}D$; then LEFT JOIN allows referencing class relation be joined with referenced class relations on equality of the reference attribute ref of the referencing class relation with OID attribute of the referenced class relations. Also attributes a.. of referenced class relations are renamed to fit post-path expressions given in the O-view signature.

**GET** ($R_D$ LEFT JOIN$_{ref=OID}$ ($R_{refD...}$ RENAME a.. AS ref.a)) [OID$_D$, ref.a..]

This operation can be nested; reference expressions of any length may be calculated in O-view attributes.

*Example.* Next O-view signature

**DOCS(.DocN, .Cntr.Name)**

is calculated as

**GET** ($R_{DOCS}$ LEFT JOIN$_{Cntr=OID}$ ($R_{CONTRACTOR}$ RENAME Name AS Cntr.Name))
[OID$_{DOCS}$, DocN, Cntr.Name]

#### 4.3.3.4 On-Reference JOIN in O-View Names

Let us name any expression defining a set of objects as group reference expression. For example the path D.ref defines set of objects of class $_{ref}$D referenced by all existing objects of class D. So, this path is group reference expression. Such group references are unary relations with OID attribute. They are calculated as O-views D(.ref) with single reference attribute which then are renamed to OID

**GET** $R_D$[ref] RENAME ref AS OID

Any path ended with reference is group reference expression.

Suppose the name of O-vies is group reference expression D.ref; then the next JOIN is used to restrict output referenced relations $R_{refD...}$ with OID given in group reference.

**GET** (D.ref JOIN$_{OID}$ $R_{refD...}$) [OID, …] , where D.ref is calculated as described above.

This operation can be nested too; reference expressions of any length may be calculated in O-view names.

*Example*. Next O-view signature

```
DOCS.Cntr( .Name, .ID)
```

is calculated as

**GET** (DOCS.Cntr JOIN$_{OID}$ $R_{CONTRACTORS}$) [OID, Name, ID]

, where DOCS.Cntr is calculated as $R_{DOCS}$[Cntr] RENAME Cntr AS OID.

#### 4.3.3.5 Object Selection Expression

Object selection expression is other way to define a set of objects and can be considered as group reference expression too.

name$_D$ < condition_list >

Its result is unary relation contained a set of OIDs of objects which belong to objects subset defined by name$_D$ of class or reference and satisfy to condition_list.

Condition_list can include a number of conditions listed by meant of commas "**,**" which means low-priority operation interrows_AND. Each of the conditions is logical expression available in WHERE operation. Its operands can be any post-paths $_D$cont of name name$_D$.

expr ($_D$cont$_1$, $_D$cont$_2$ …)

A set of object satisfied to this condition is calculated as

**GET** (D($_D$cont$_1$, $_D$cont$_2$ …) WHERE <u>expr ($_D$cont$_1$, $_D$cont$_2$ …)</u>) [OID]

Here the part D($_D$cont$_1$, $_D$cont$_2$ …) is implicit O-view which is calculated using operations given above. The result is OIDs of objects satisfying to condition. It's interesting that recording of the condition stays the same (underlined in calculation expression).

Interrows_AND operation is calculated by relational target machine as INTERSEPT of the sets of OID which are results of calculating of each of conditions bound by commas. So the next expression

expr($_D$cont$_1$ …), expr$_2$($_D$cont$_2$ …)

is calculated as

**GET** (D($_D$cont$_1$ …]) WHERE expr$_1$($_D$cont$_1$ …)) [OID]
    INTERSEPT
    (D($_D$cont$_2$ …]) WHERE expr$_2$($_D$cont$_2$ …)) [OID]

*Example*. Next O-view signature

```
DOCS<.Items.Art = "Tie", .Items.Art = "Axe">(.DocN)
```

is calculated as

**GET** (
((R$_{DOCS.Items}$ RENAME Art AS Items.Art) WHERE Items.Art = "Tie") [OID]
INTERSEPT
((R$_{DOCS.Items}$ RENAME Art AS Items.Art) WHERE Items.Art = "Axe") [OID]
) JOIN$_{OID}$
R$_{DOCS}$[OID, DocN]

## 4.4 Implementation Command

General syntax of the implementation command is

ALTER D REALIZE name AS expression

, where D is name of some class, name is a name of its component or method, expression describes how the component or the method is implemented. The command describes an implementation and binds the implementing expression with name of component or method.

In this part translation of the implementing expressions are discussed.

Class components can be stored or calculated. Stored component is implemented with command.

ALTER D REALIZE C AS STORED

, where C is the component.

Calculated components are implemented with command

ALTER D REALIZE C AS RValue

, where RValue is any expressions returning value which is not complex than a value of relation.

The methods of class are implemented by procedures using the command

ALTER D REALIZE M(…) AS begin … end,

, where M(…) is the method signature, begin … end is implementing procedure.

### 4.4.1  Group Execution of Operations and Methods

Any operations on objects and object methods do not need iterators to be executed for group of objects. Let's illustrate this ability with simple example. Values of two scalar components **a** and **b** are summed in some implementing expression as **a + b**.

Translation of this operation has to include access operation to values contained in correspondent class relations R$_D$…. As a result the translation roughly looks like the next expression

**GET** (R$_D$ WHERE OID = this)[a]  + (R$_D$ WHERE OID = this)[b]

, where this contains OID of object which the implementation are executed for.

Let us note that the access operation can be "factorized"

**GET** (R$_D$ WHERE OID = this)[a+b]

It's obvious that the last expression can be executed for group of objects which is defined by group reference these as

**GET** (R$_D$ JOIN$_{OID}$ these)[OID, a+b]

The result is relation containing all sum values for each of objects referenced by these. Also it contains attribute OID; so it is structurally unified to class relations and can be used in next operations. It's interesting that the recording of the scalar operation stays the same (underlined).

This ability seems to be universal and can be formulated as the next Translation Thesis:

Any procedure p(…) on single object of class D can be translated in such procedure p'(these, …) on the class relations R$_D$… that result of single execution of p' is equal to execution of p for each of objects referenced by any group reference these.

The same statement for single operations on objects can be considered as special case of the Translation Thesis.

(Remark. Further the objects of class D referenced by group reference **these** are named as **these** objects. The group reference **these** is hidden from user. Its value is the result of calculation of some reference expression THESE given in source code.)

The Thesis means e.g. that the expression EXEC THESE.M(…), where M(…) is the method implemented in class D by some procedure p(…), is executed in two next steps:
1. Group reference **these** is calculated according to reference expression THESE,
2. Translation **p'** of source p(…) is executed taking **these** as input parameter.

**EXEC** p'($_{in}$these, …)

At that all data existing in class relation R are changed as if the source procedure p(…) was executed for each of objects defined by source expression THESE.

The Translation Thesis is discussed in next paragraphs. The discussion can be considered as description of principles of translation of operations and procedures.

### 4.4.1.1 Translation of Expressions Defining Parameters and Local Variables

Let us note that logic of representing component C values in form of class relation $R_{D…}$ (see 4.2.2) can be also applied to parameters and to local variables of procedures. Scalar parameters $_{sc}par_1$, $_{sc}par_2$, … of procedure p can be represented as parameter relation $R_{par}(OID, {}_{sc}par_1, {}_{sc}par_2,…)$, which is parameter of **p'** and must be created in code which executes **p'**. This relation contains scalar parameter values for each of **these** objects.

**GET** $R_{par}$ [OID]   ->   $_{in}$these

If procedure p take no parameter, the parameter relation $R_{par}$ contains only attribute OID.

The same logic is true for local variables which can be represented in relation $R_{local}$. This relations must be created inside of **p'**.

(Remark. For sake of simplicity we consider scalars only but generally the parameters and locals can be values which are not more complex than relation.)

Translation of expressions describing parameters and local variables is similar to translation of class components specification. The result of the translation is temporary real relational variable.

In fact, the parameters and local variables are certainly defined inside procedure as well as other class components are. They can be treated as non-persistent class components with different lifetime. So, when procedures are discussed, they will be considered as elements of set $C_i$ of class components. Accordingly parameters relation $R_{par}$ and local variables relation $R_{local}$ will be considered as elements of set $R_{D…}$ of class relations further.

### 4.4.1.2 Translation of RValue expressions.

Let us consider some operation **RValue** on values C, which describe the state of some class D object. Let us note that the values C are not only values of object component $C_i$. The name of class components $C_i$ available in scope of some class D can be considered as a short record of expression "this.$C_i$", where **this** is a name of implicit self-reference variable (i.e. special case of group reference expression). With such consideration it is possible to say that in scope of class D a set of local O-views C is defined, whose full signatures includes the name of self-reference variable "**this**" and (then) the name of class component "$C_i$" as first two items. In a simplest case for scalar object components $_{sc}C_i$ the full signature of such local O-view is **this**($_{sc}C_i$). More complex cases are possible if path "this.$C_i$" allows post-paths. In these cases expression "this.$C_i$" appears in the beginning of local O-view signature as a name of the O-view or as a start part of this name.

For each of local O-view C a corresponding O-view $R_C$ exist such as

**GET** ($R_C$ WHERE OID = this)   –>   C

Signature of O-view $R_C$ differs from the signature of corresponding local O-view C in first element only; it begins with name of class D instead of name of self-reference variable **this**.

Let us consider some operation **RValue** on object views C, which calculates a relational value

f($C_1$, $C_2$, …)   –>   $C_n$                                                                                          (1)

Here $C_n$ is a result of operation, f is composition of primitive relational operation $_{prim}op$ on object views $C_1$, $C_2$, … available in scope of class D

$_{prim}op_1(C_1,\ _{prim}op_2(C_2, …\ (…)))$

At that, the primitive relational operations $_{prim}op$ can contain any scalar operations on attributes of operands.

For each of primitive relational operations $_{prim}op$ the next is true.

− Union $C_1 \cup C_2$ is equal to
  **GET** (($R_{C1} \cup R_{C2}$) WHERE OID = this)

− Difference $C_1 - C_2$ is equal to
  **GET** (($R_{C1} - R_{C2}$) WHERE OID = this)

− Cartesian product $C_1 \times C_2$ is equal to
  **GET** (($R_{C1}$ JOIN$_{OID}$ $R_{C2}$ WHERE OID = this)

− Selection C WHERE condition is equal to
  **GET** (($R_C$ WHERE condition) WHERE OID = this)

− Projection $C[a_1, a_2, ...]$ is equal to
  **GET** (($R_C[OID, a_1, a_2, ...]$) WHERE OID = this)

At that, all possible in $_{prim}op$ scalar operations on attributes of object views C are applied to corresponding relations R attributes without changes.

So, all primitive operations $_{prim}op(C_1 …) \to C_{res}$ ($C_{res}$ is result of the operation) is deduced to expression

**GET** (op'($R_{C1}…$) WHERE OID = this)   –>   $C_{res}$

(2)

Let's note that logic of representing component C values in form of class relation $R_{D…}$ (see 4.2.2) can be applied to values $C_{res}$. Result relation $R_{Cres}$ unites results of $_{prim}op(C_1 …)$ executed for all class D objects.

**GET** ($R_{Cres}$ WHERE OID = this)   –>   $C_{res}$

(3)

Comparison of (2) and (3) gives that any primitive operation $_{prim}op$ on C can be deduces to operation op' on $R_C$ which produces result relation $R_{Cres}$.

**GET** op'($R_{C1} …$)   –>   $R_{Cres}$

Relational algebra closure allows any result relation $R_{res}$ be used as operand of other operation

**GET** op'(… $R_{Cres}$ …)   –>   $R_{Cres+1}$

Thus, for any operation (1) such operation f' on $R_i$ exists

**GET** f'($R_{C1}, R_{C2}, …$)   –>   $R_{Cn}$

that its result $R_n$ unites all results $C_n$ of source operation f executed in each of class D objects

**GET** ($R_{Cn}$ WHERE OID = this)   –>   $C_n$

Let's name f' as translation of source operation f. Result of operation

**GET** f'($R_{C1}, R_{C2}, …$) JOIN$_{OID}$ these   –>   $_{these}R_{Cn}$

is relation $_{these}R_{Cn}$ which unites result of source operation f executed in each of these objects.

#### 4.4.1.3    Translation of Procedures

Let us consider a procedure p as algorithmic sequence of assignment operations

$C_n := f(C_1, C_2, \ldots)$

and calls of methods

EXEC M ($par_1$, $par_2$, …), where $par_i$ – parameters of the method.

Assignment operation

$C_n := f(C_1, C_2, \ldots)$

is translated in

**SET** $R_n$ :=
$f'(R_1, R_2, \ldots)$ JOIN$_{OID}$ **these**
UNION
($R_n$ WHERE JOIN$_{OID}$ ($R_n$[OID] MINUS **these**))

, where part $f'(R_1, R_2, \ldots)$ JOIN$_{OID}$ **these** is result of f' execution on data of **these** objects, part ($R_n$ WHERE JOIN$_{OID}$ ($R_n$[OID] MINUS **these**)) is unchanged part of $R_n$, containing data of not **these** objects.

This translation updates relation $R_n$ by means of replacing tuples containing data of **these** objects with new ones. Further it is written as

**REPLACE** $R_n$ with ($f'(R_1, R_2, \ldots)$ JOIN **these**)

Call of method

EXEC M ($par_1$, $par_2$, …)

is translated in call of its translation

**EXEC** p'($R_{par}$)

, where $R_{par}$(OID, $par_1$, $par_2$,…) is parameter relation such as $R_{par}$[OID] -> **these**.

Linear sequence of described command *Sqns* is translated in the same sequence of their translations ***Sqns'*(these)**. For example the next linear sequence of commands

$C_n := f_1(C_1, C_2, \ldots)$
exec p ($par_1$, $par_2$, …)
$C_{n+1} := f_2(C_1, C_2, \ldots)$

is translated in the same sequence of their translations

**REPLACE** $R_n$ with ($f'_1(R_1, R_2, \ldots)$ JOIN **these**)
**EXEC** p'($R_{par}$)
**REPLACE** $R_{n+1}$ with ($f'_2(R_1, R_2, \ldots)$ JOIN **these**)

To translate common algorithm structures **if…** , **while…** [Bohm and Jacopini, 1966] a number of group reference variables $_{br}$**these** are used which are local in p'(). Each $_{br}$**these** variable corresponds to one of algorithm branch. Variables $_{br}$**these** are empty when p' is starting. Then their values are changed according to the algorithm and to the used condition. For all $_{br}$**these**$_i$ used in p'() the next two rules are always true

$\bigcup {}_{br}\text{these}_i = {}_{in}\text{these}$

$_{br}\text{these}_i \cap {}_{br}\text{these}_j = \varnothing$ , $i \neq j$

Algorithm structure

**If** condition **then** *Sqns$_1$* ;

is translated into next sequence of command of relational target mashine

**SET** $_{br}$these$_{True}$ := $_{br}$these<condition'>
**SET** $_{br}$these := $_{br}$these - $_{br}$these$_{True}$
**if** COUNT($_{br}$these$_{True}$) **then** *Sqns$_1$'*($_{br}$these$_{True}$)
**SET** $_{br}$these := $_{br}$these $\cup$ $_{br}$these$_{True}$
**SET** $_{br}$these$_{True}$ := $\varnothing$

Here the group reference $_{br}$these corresponds to algorithm branch which precedes the **if** operator. The group reference $_{br}$these$_{True}$ contains identifiers OID of $_{br}$these objects which satisfy to **condition** given in source code (here an object selection expression is used). COUNT ($_{br}$these$_{True}$) is operation returning number of elements in group reference $_{br}$these$_{True}$ (i.e. cardinality of $_{br}$these$_{True}$). As a result a part of algorithm after **then** operator is executed only for objects which satisfy the given **conditions.**

Algorithm structure

**while** condition
   *Sqns$_1$*;

is translated as

**SET** $_{br}$these$_{True}$ := $_{br}$these<condition'>
**SET** $_{br}$these := $_{br}$these - $_{br}$these$_{True}$
**while** COUNT($_{br}$these$_{True}$)>0
**begin**
  *Sqns$_1$'*($_{br}$these$_{True}$)
  **SET** $_{br}$these := $_{br}$these UNION ($_{br}$these$_{True}$ - $_{br}$these$_{True}$<condition'> )
  **SET** $_{br}$these$_{True}$ := $_{br}$these$_{True}$<condition'>
**end**

Here the loop will be continuing while objects satisfying to condition exist in $_{br}$these$_{True}$ subset**.**

Let us also note that any algorithm branches can contain statements **return** which cause execution to leave the current procedure and resume at the point in the code immediately after where the procedure was called. This statement is translated as

**SET** $_{out}$these := $_{out}$these UNION $_{br}$these
**SET** $_{br}$these := ∅

This translation does not leave the current procedure but empties current subset $_{br}$these and adds its value to special variables $_{out}$these which collects all the subsets which goes through branches containing **return** statements. All $_{out}$these objects stay intact while other algorithm branches are executed for other objects.

Thus, any expressions and procedures which implement class components and methods can be translated in operations and procedures of relational target machine, which manipulates the data of group of objects at once.

### 4.4.2 Translation of STORED Implementation Expression

Stored components are implemented in relational target machine with real relational variables.

**CREATE** $_{real}$R$_{D...}$ ( OID, … )…;

Schemas of the real variables are equal to schemas of corresponding class relations R$_{D...}$. Real variables which implements the class scalar relations R$_D$ (see 4.5.1) exist always.

### 4.5 Implementation Binding

In this part the ALTER … REALIZE… command are discussed as a command which binds component or method with implementing expression.

Implementations of components and methods can be changed during inheritance. So when component is accessed or method is called it has to be bound with proper implementation. In group operation a number of implementation can be used.

Considering the binding expression translation we will use the next case. Suppose class D was created. Its component C is implemented as stored. Method M(…) is implemented with procedure p$_1$

CLASS D(…C, M(…), …);
ALTER D REALIZE C AS STORED;
ALTER D REALIZE M(…) AS p$_1$;

Child class $_{sub}D$ is created. In it the inherited component C is re-implemented as calculated by operation f(…) and the inherited method M is re-implemented with procedure $p_2$

CLASS $_{sub}D$ EXTEND D …;
ALTER $_{sub}D$ REALIZE C AS f(…);
ALTER $_{sub}D$ REALIZE M(…) AS $p_2$;

### 4.5.1 Component Binding and Class Relation Calculation

Components are bound during class relations calculation. The expression which is used to calculate virtual class relation $R_{D…}$ is the binding one.

Binding expression for complex components is UNION of results given by all implementations. If component C of class D is complex ones then corresponded class relation $R_{D.C}$ unites translations of implementation expressions which are set in class D and in child class $_{sub}D$. The next binding expression is used for the calculation.

**GREATE** $R_{D.C}(…)…$ AS
$_{real}R_{D.C}$
  UNION
$f_1'(…)$ ,

where $_{real}R_{D.C}$ is implementation as stored component (as it is set in class D), $f_1'(…)$ is translation of realization f(…) set in child class $_{sub}D$.

Binding expression for simple components is replaced the attribute of stored scalar class relation with calculated value. If component C of class D is complex ones then corresponded class relation $R_{D.C}$ is calculated with the next expression

**GREATE** $R_D$ (…)… AS
$_{real}R_D$ LEFT JOIN$_{OID}$ ($f_1'(…)$ [OID, calcC]) [OID, …, SUBST(C, $_{calc}$C), …]

, where $_{real}R_D$ is stored scalar class relation, $f_1'(…)$ is translation of realization f(…) set in child class $_{sub}D$, SUBST(C, $_{calc}$C) is operator substituting attribute of simple component for calculated value.

It's clear that any changes in component implementation have to result in the changes of the binding expression. Thus translation of source command

ALTER D REALIZE C AS …

has to re-create the binding expression executed by target machine to calculate class relations.

### 4.5.2 Method Binding

A number of implementing procedures can be executed if method is called for group of **these** objects. In this case binding expression is a procedure which calls all the method implementations. At that for each of the implementations the corresponding subset of **these** objects has to be determined.

Method call for group of objects

exec THESE.M(…)

means the execution of the next stored transaction

**TRANS** D.M'(these, …) AS
**begin**
  **EXEC** $p'_1$(these INTERSEPT scope($p_1$), …);
  **EXEC** $p'_2$(these INTERSEPT scope($p_2$), …);
**end**

,where $p'_1$ and $p'_1$ are translation of implementing procedures $p_1$ and $p_2$, scope(p) is a group reference on objects which the implementation p is bound with.

It's clear that any changes in method implementation have to result in the changes of the binding expression. Thus translation of source command

ALTER D REALIZE M(…) AS …

has to re-create the binding procedure D.M'(**these**, …) which is ran in target machine to execute proper implementations of the method.

## 4.6 Object Creation Command

Object is created using the instruction NEW.

NEW D constructing_expression

Let us note that it can be used both as in-procedure instruction and as a declarative language command. In the last case it doesn't return reference on new objects. Any objects are element of classes and can be accessed using data access commands by values of its components which could be set by constructing_expression (any command sequence or procedure defined in context of class D).

### 4.6.1 Translation of Object Creation Command

During object creation the relational target machine performs the next steps
1. New unique OID values is generated
2. Tuples with new OIDs is added into the stored class scalar relation $_{real}R_D$. For object of child classes all such relations of parent classes have to be added with the tuples too.
3. Translated construction_expression' is performed using new OID values as these group reference

## 5. COMPATIBILITY OF CLASES AND RELATIONS

Because of using the relational data model as format basis and the relational programmable system as target machine the offered approach doesn't impede collateral usage of both classes and relations which are native data structure for the target machine. There are a three ways of how classes and relations can be co-used. Let us consider class D and relation T.

1) Relations can be defined on a set of domains which is extended with new reference types. Suppose relation T contains attribute ref of reference domain D.

   CLASS D(C:…);

   **CREATE** T(i:INTEGER, ref:D);

   Then any of the class D post-paths can be used when the relation T is accessed.

   T(i, ref.C)

   The last relation is calculated in relational target machine by meant of JOIN on relation attributes ref. Also attributes are renamed to fit path expressions given as attribute name.

   **GET** (T  LEFT JOIN$_{ref = OID}$ (R$_D$  RENAME C AS ref.C)) [i, ref.C]

2) Mutual constraints (foreign key) can be set for any class D and any relation T. Such constraints implemented in relational target machine as the ones set on the relation T and corresponded class relation R$_{D…}$.

3) Because data of classes presented in form of relations (object views) they can be combined with the explicit relations in data access operations (e.g. in ad-hoc queries).

## 6. CONCLUSION

We introduced base principles of object-oriented translation for programmable relational system. Offered approach allows formal-based evolution of existing relational systems towards the systems which can be described as independent object-oriented environments for active persistent manageable model of problem domain. At that possibilities and features of existing relational systems (e.g. transaction, access protocols etc.) stay the same so current population of applications used the systems may stay intact.

In conclusion let us recall the idea of data independence defined by E.F.Codd as "the independence of application program and terminal activities from growth in data types and changes in data representation". The term "data" can be treated very widely. It means both values describing the state of objects of problem domain and functionality which reflects the objects mutability and their dependences. All the data form a model of problem domain and in this case the data independence means independence of the model.

## ACKNOWLEDGEMENT

The author thanks Sergey D. Kuznetsov for numerous works which was fertile ground for offered suggestions.